\documentclass{ws-ijmpe}
\usepackage[super,compress]{cite}

\def\half{{\textstyle{1\over 2}}}

\begin{document}

\markboth{Jun He and Pei-Liang L\"u}{The Octet Meson and Octet Baryon Interaction with strangeness and the $\Lambda(1405)$}

\catchline{}{}{}{}{}

\title{The Octet Meson and Octet Baryon Interaction with strangeness and the $\Lambda(1405)$}

\author{Jun He$^{1,3,4}$ and Pei-Liang L\"u$^{1,2}$}

\address{$^1$ Theoretical Physics Division, Institute of Modern Physics, Chinese Academy of Sciences,
Lanzhou 730000, China\\
$^2$University of Chinese Academy of Sciences, Beijing 100049, China \\
$^3$ Research Center for Hadron and CSR Physics,
Lanzhou University and Institute of Modern Physics of CAS, Lanzhou 730000, China\\
$^4$ State Key Laboratory of Theoretical Physics, Institute of
Theoretical Physics, Chinese Academy of Sciences, Beijing  100190,China\\
junhe@impcas.ac.cn}

\maketitle

\begin{history}
\received{Day Month Year}
\revised{Day Month Year}
\end{history}

\begin{abstract}
The octet meson and baryon interaction with strangeness $S=-1$ is
studied fully relativistically with chiral Lagrangian. In this work, a Bethe-Salpeter
equation approach with spectator quasipotential approximation is applied to study the
reactions $K^-p\to MB$ with $MB=K^-p,\bar{K}^0n,
\pi^-\Sigma^+,\pi^0\Sigma^0, \pi^+\Sigma^-$ and $\eta \Lambda$ with all possible partial waves and theoretical results are comparable with
experimental data.  It is found that the Weinberg-Tomozawa potential derived from the
lowest order chiral Lagrangian only provides the contributions from
partial waves with spin-parities $J^P=1/2^+$ and $1/2^-$. Two-pole structure of the $\Lambda(1405)$ is confirmed with poles at $1383+99 i$ and $1423+14i$ MeV. The lower and higher poles originate
from $\Sigma \pi$ interaction as a resonance and $\bar{K}N$
interaction as a bound state, respectively.

\end{abstract}


\ccode{PACS numbers:14.20.Jn, 13.75.Jz, 11.10.St, 12.39.Fe}


\section{Introduction}

Among the observed excited hyperons, the $\Lambda(1405)$ with spin parity $J^P=1/2^-$ and isospin $I=0$ attracts peculiar attention \cite{Moriya:2014kpv}. The
existence of $\Lambda(1405)$ was theoretically predicted in 1959 by
Dalitz and Tuan \cite{Dalitz:1959dn,Dalitz:1960du}, and confirmed in
the experiment as early as 1961 in the $\pi\Sigma$ invariant mass spectrum~\cite{Alston:1961zzd}. After a half century,
$\Lambda(1405)$ has been well established experimentally and is
currently listed as a four-star resonance in the table of the Particle
Data Group (PDG) \cite{Agashe:2014kda}.  However, there is still no
universal agreement on the nature of this state.

As the first excited $\Lambda$ resonance, the $\Lambda(1405)$ should
be assigned as a P-wave baryon in the constituent quark model
\cite{Isgur:1978xj,Capstick:1986bm}. However, althrough the  constituent
quark model provides a general scheme to describe the non-strange low-mass
baryons, it has difficulty in computing the $\Lambda(1405)$
mass. The theoretical mass, about 1500 MeV, is about one hundred MeV
higher than the observed one. It is
even lower than the mass of its nonstrange counterpart $N(1535)$,
which is very strange because in the constituent quark model strange quark
is about one third heavier than nonstrange quarks. Due to such
difficulties it is popular to adopt an interpretation beyond the three
quark picture for the $\Lambda(1405)$ in the literature.

Within the framework of the constituent quark model, a proposal was made
to solve the reverse mass order  problem for $\Lambda(1405)$ and
$N(1535)$  by introducing  five-quark component \cite{Zou:2006tw}.
Besides nonstrange quarks, the suggested five-quark component in a
$N(1535)$ carries a strange-quark pair while in a $\Lambda(1405)$
the five-quark component carries only one strange quark, which naturally leads to a conclusion in such picture
that the $N(1535)$ is heavier than the $\Lambda(1405)$.  A
more popular interpretation of the $\Lambda(1405)$ is dynamically
generated state with a two-pole structure within chiral unitary
approach which combines the low energy interaction governed by chiral symmetry and the
unitarity condition for the coupled-channel scattering amplitude
\cite{Oller:2000fj,Oset:1997it, Jido2003,Hyodo:2011ur}.  Another theory
describes the state as a quasibound state of $N\bar{K}$ embedded in a
$\Sigma\pi$ continuum with a one-pole structure
\cite{Akaishi:2010wt}.  A recent lattice QCD
investigation also supports that the structure of the $\Lambda(1405)$
is dominated by a molecular bound state of an anti-kaon and a nucleon
\cite{Hall:2014uca}.

In the chiral unitary approach \cite{Krippa:1998us,Oller:2000fj,Oset:1997it, Jido2003,Hyodo:2011ur}, an on-shell factorization is
adopted. In such approach,  in potential both baryon and meson are put on shell, which makes it possible to factorize the potential out of the integral. It leaves a loop function where a sum over all intermediate (off-shell) states is performed. After such factorization the Bethe-Salpeter equation (BSE) becomes an algebra equation instead of an integral equation. In this work, we follow another way to solve the Beth-Salpeter equation. In the literature, a BSE approach with a spectator
quasipotential approximation, which is fully relativistic, has been developed to study the deuteron  \cite{Gross:2010qm} and extended to study the hadronic
molecular state recently, such as the $Y(4274)$, the $\Sigma_c(3250)$ and the $N(1875)$~\cite{He:2011ed,He:2012zd,He:2013oma, He:2015yva, He:2014nya}. In Ref. \refcite{He:2015mja}, it was extend to study the scattering amplitude to search for the pole in the second Riemann sheet which is related to the $Z_c(3900)$ observed at BESIII as a resonance instead of as a bound state.  In this work, we will study  octet meson and baryon interaction with strangeness $S=-1$ and the relevant $\Lambda(1405)$ in the
BSE approach with the quasipotential approximation.

The paper is organized as follows. In next section, the potential  is
derived with the  help of the chiral Lagrangian.  In section 3 we
develop a theoretical frame to study the octet meson and baryon interaction by solving the coupled-channel BSE  with the quasipotential approximation. The numerical results are
given in Section 4. In the last section, a summary is given.

\section{The chiral Lagrangian and potential}

The potential from the chiral Lagrangian has been well discussed in the
literature \cite{Oller:2000fj,Oset:1997it, Jido2003,Hyodo:2011ur}. At the
lowest order in the chiral expansion, besides the Weinberg-Tomozawa term
there are also the $s$ and $u$-channel diagrams
involving the coupling of the meson-baryon channel to an intermediate
baryon state.  However, the contribution of these terms is very
moderate \cite{Oller:2000fj}. Although they were shown to
be helpful in reproducing more physical values of the subtracting constants in
some cases, they do not influence significantly
the quality of the fits \cite{Ikeda:2012au}. We will neglect these terms in the present
study. And from a recent study~\cite{Feijoo:2015yja} the
next-to-leading order is only sensitive to the $\bar{K}N\to K\Xi$
reaction, which is not focused on in this work. Hence, only
the Weinberg-Tomozawa term will be included in the current
work.

The lowest order chiral
Lagrangian, coupling the octet of pseudoscalar mesons to the octet of
$1/2^+$ baryons, is written as
\cite{Gasser:1987rb,Pich:1995bw,Ecker:1994gg,Bernard:1995dp}
\begin{eqnarray}
{\cal L}_1^{(B)}=\frac{1}{4f^2}\langle \bar{B}i\gamma^\mu
[\mathbb{P}\partial_\mu \mathbb{P}-\partial_\mu \mathbb{P} \mathbb{P},B]\rangle,
\end{eqnarray}
where $B$ and $\mathbb{P}$ stand for  baryon and pseudoscalar meson fields, and the constant $f$ is the meson decay constant.

The potential is easily evaluated  and  given by
\begin{eqnarray}
i{\cal V}^{kl}_{\lambda\lambda'}=-C^{kl}\frac{1}{4f^2} \bar{u}(p)\gamma^\mu u(p') (k_\mu+k'_\mu),
\end{eqnarray}
where $u(p')$ and $\bar{u}(p)$ are the Dirac spinors and $k^{(')}$ and $p^{(')}$ are the momenta of the outgoing (incoming) meson and baryon.
The coefficient $C^{kl}$ for channel $l$ to channel $k$ is fixed by chiral
symmetry, and the explicit value was given in Ref.~\refcite{Oset:1997it}.
In this work the $K^-p\to MB$ reactions will be investigated and ten interaction channels $K^-p$, $\bar{K}^0n$, $\Sigma^+\pi^-$, $\Sigma^-\pi^+$,
$\Sigma^0\pi^0$, $\Lambda\eta$, $\Lambda\pi^0$, $\Sigma^0\eta$,
$K^+\Xi^-$ and $K^-\Xi^+$ will be  included in a coupled-channel calculation.
The $\Lambda(1405)$ is an isoscalar state,
so the isospin zero state will be constructed as in Ref. \refcite{Oset:1997it} to
study the relation between the interactions and the $\Lambda(1405)$,
in which four interaction channels $\bar{K}N$, $\pi\Sigma$,
$\eta\Lambda$ and $K\Xi$ are included. The corresponding coefficient $D^{kl}$ for a channel with fixed isospin was also given in Ref.~\refcite{Oset:1997it}.

Since the potential kernel will be used in the BSE, both initial and final particles are put off shell firstly. With such requirement the potential can be expressed explicitly as
\begin{eqnarray}
i{\cal V}^{kl}_{\lambda\lambda'}&=&-C^{kl}\frac{1}{4f^2} \bar{u}(p')\gamma^\mu u(p) (k_\mu+k'_\mu)\nonumber\\
&=&\sqrt{\frac{p^0+\tilde{M}}{2\tilde{M}}}\sqrt{\frac{p'^0+\tilde{M}'}{2\tilde{M}}}\phi^\dag_\lambda\left\{k^0+k'^0+
\big[{\bm p}^2(p'^0+\tilde{M}')+{\bm p}'^2(p^0+\tilde{M})\right.\nonumber\\
&+&\left.(2W+\tilde{M}+\tilde{M}')({\bm p}\cdot{\bm p}'+i{\bm \sigma}\cdot{\bm p}\times{\bm p}')\big]\frac{1}{(p'^0+\tilde{M}')(p^0+\tilde{M})}\right\}\phi_{\lambda'},\quad\quad
\end{eqnarray}
where  the $W$, $M^{(')}$ and ${\bm p}^{(')}$ are total energy of system and the mass and momentum  of the outgoing (incoming)
baryons  in the center of mass frame. The $\tilde{M}=\sqrt{p^2}$ with $p^2$  being the square of four momentum of the baryon.
The $\phi_\lambda$ and $\phi_{\lambda'}$ for the initial and final baryons with helicities $\lambda$ and $\lambda'$ are defined as
\begin{eqnarray}
\phi^T_{+1/2}&=&(\cos(\theta/2),\sin(\theta/2)e^{i\varphi}),\nonumber\\
\phi^T_{-1/2}&=&(-\sin(\theta/2)e^{-i\varphi},\cos(\theta/2)),
\end{eqnarray}
where $\theta$ and $\varphi$ are angles of the baryon momentum.

To reach the popular form of potential as given in Ref. \refcite{Hyodo:2011ur} where on-shell factorization was adopted, we do not need to adopt the onshellness of both baryon and meson in the potential. Here only the baryon is put on-shell, which means $\tilde{M}=M$ and
$E=\sqrt{M^2+{\bm p}^2}$. The potential is rewritten as \cite{Hyodo:2011ur},
\begin{eqnarray}
i{\cal V}^{kl}_{\lambda\lambda'}
&=&-\frac{C^{kl}}{4f^2}\sqrt{\frac{E+M}{2M}}\sqrt{\frac{E'+M'}{2M}}\nonumber\\
&\cdot&\phi^\dag_\lambda\big[2W-M-M'+
\frac{(2W+M+M')({\bm p}\cdot{\bm p}'+i{\bm \sigma}\cdot{\bm p}\times{\bm p}')}{(E+M)(E'+M')}\big]\phi_{\lambda'}.
\end{eqnarray}
The
$\phi_\lambda$ used here is dependent on the angles $\theta$ and
$\varphi$.  Without loss of generality, the momenta can be chosen as
$p'=(E',0,0,{\rm p})$ and $p=(E,{\rm p}\sin\theta,0,{\rm
p}\cos\theta)$ with ${\rm p}=|{\bm p}|$ in order to avoid confusion with the
four-momentum $p$.

The potential can be rewritten further as
\begin{eqnarray}
i{\cal V}^{kl}_{\lambda\lambda'}
&=&-\frac{C^{kl}}{4f^2}\sqrt{\frac{E+M}{2M}}\sqrt{\frac{E'+M'}{2M}}\nonumber\\
&\cdot&\left[(2W-M-M')+\frac{(2W+M+M'){\rm p}{\rm p}'}{(E+M)(E'+M')}(-1)^{\lambda'-\lambda}\right]d^\half_{\lambda'\lambda}(\theta).  \label{Eq: potential with d}
\end{eqnarray}
\normalsize
We will see soon that the potential contributes  only to $J=1/2$ partial
waves because it is proportional to a Wigner $d$-matrix
$d^\half_{\lambda'\lambda}(\theta)$.

\section{The coupled-channel BSE for scattering amplitude}

In Ref. \refcite{He:2014nya}, the BSE of the vertex function are adopted to study the bound state.
In this work, we deal with the BSE for the scattering amplitude ${\cal
M}$ directly, which is written as~\cite{He:2015mja}
\begin{eqnarray}
	{\cal M}={\cal V}+{\cal V}G{\cal M},
\end{eqnarray}
where the ${\cal V}$ is the potential kernel and the $G$ is the propagator
for two constituent particles.

To avoid the difficulties in the numerical solution, the four-dimensional BSE is usually reduced to a three-dimensional equation.  There
exist many methods to make such three-dimensional reduction, which include the $K$
matrix approximation, the Blankenbecler-Sugar approximation and the covariant
spectator theory~\cite{Nieuwenhuis:1996mc,Blankenbecler:1965gx,Logunov:1963yc,Shklyar:2004ba}.  In this work, we adopt the covariant spectator
theory.

With the help of onshellness of
the heavier constituent 2, baryon, the propagator can be
\begin{eqnarray}
G(p,P)&=&-\frac{\sum_{\lambda}u_{\lambda}(p)\bar{u}_{\lambda}(p)
}{(p^2-M^2)((P-p)^2-m^2)}
\equiv \tilde{G}(p)\sum_{\lambda}u_{\lambda}(p)\bar{u}_{\lambda}(p),
\end{eqnarray}
where $u_{\lambda}(p)$ is the spinor with
helicity $\lambda$, and $M$ and $m$ are the masses of baryon and meson.
After spinor is multiplied  on both sides, the BSE results
\begin{eqnarray}
{\cal M}^{kl}_{\lambda,\lambda'}&=&{\cal
V}^{kl}_{\lambda,\lambda'}+\sum_{n,\lambda''}{\cal V}^{kn}_{\lambda\lambda''}
G^{n}_0{\cal M}^{nl}_{\lambda'',\lambda'},
\end{eqnarray}
where $k,l$ or $n$ is for different channel $MB=K^-p,\bar{K}^0n,
\pi^-\Sigma^+,\pi^0\Sigma^0, \pi^+\Sigma^-$ or $\eta \Lambda$ in this work and will be omitted if not
necessary.
The remnant of propagator $\tilde{G}(p)$  is written down in
the center of mass frame where $P=(W,{\bm 0})$  is
\begin{eqnarray}
	\tilde{G}({\rm p})&=&2\pi i\frac{\delta^+(p^2-M^2)}{(P-p)^2-m^2}
	=2\pi
	i\frac{\delta^+(p^0-E({\rm p}))}{2E({\rm p})[(W-E({\rm
p}))^2-\omega({\rm p})]}\\\nonumber&\equiv&2\pi i \delta^+(p^0-E({\rm p})) G_0({\rm p}),
\end{eqnarray}
where  $E({\rm p})=\sqrt{M^2+{\rm p}^2}$
and $\omega({\rm p})=\sqrt{m^2+{\rm p}^2}$.

To reduce the BSE to one-dimensional equation, we apply the
partial wave expansion as
\begin{eqnarray}
{\cal V}_{\lambda\lambda'}({\bm p},{\bm p}')\
&=&	\sum
	 _{J\lambda_R}{\frac{\sqrt{2J+1}}{4\pi}}D^{J}_{\lambda_R,\lambda}(\varphi,\theta,0){\cal
V}^J_{\lambda\lambda'}({\rm p},{\rm p}')D^{J*}_{\lambda_R,\lambda'}(\varphi',\theta',0),
\\
{\cal V}_{\lambda\lambda'}^J({\rm p},{\rm p}')&=&2\pi\int d\cos\theta d^{J}_{\lambda'\lambda}(\theta)
{\cal V}_{\lambda\lambda'}({\bm p},{\bm p}'),
\end{eqnarray}
where the momenta are chosen as $p'=(E',0,0,{\rm p})$,
and $p=(E,{\rm p}\sin\theta,0,{\rm p}\cos\theta)$.
Combined with Eq. (\ref{Eq: potential with d}), one can find that the Weinberg-Tomozawa  potential derived from the
lowest order chiral Lagrangian only provides the contributions for
partial waves with spin $J=1/2$.

Now we have the partial wave BSE as,
\begin{eqnarray}
{\cal M}^J_{\lambda\lambda'}({\rm p},{\rm p}')
&=&{\cal V}^J_{\lambda,\lambda'}({\rm p},{\rm p}')+\sum_{\lambda''}\int\frac{{\rm
p}''^2d{\rm p}''}{(2\pi)^3}
{\cal V}^J_{\lambda\lambda''}({\rm p},{\rm p}'')
G_0({\rm p}''){\cal M}^J_{\lambda''\lambda'}({\rm p}'',{\rm
p}').\quad\quad
\end{eqnarray}

The partial wave BSE with fixed parity is of a form,
\begin{eqnarray}
{\cal M}^{J^P}_{\lambda\lambda'}={\cal V}^{J^P}_{\lambda\lambda'}+\sum_{\lambda''>0}
{\cal V}^{J^P}_{\lambda\lambda''}G{\cal M}^{J^P}_{\lambda''\lambda'},
\end{eqnarray}
where the amplitude with fixed spin parky is defined as \begin{eqnarray}
{\cal M}^{J^P}_{\lambda\lambda'}={\cal M}^{J}_{\lambda\lambda'}+\eta {\cal M}^J_{\lambda,-\lambda'},
\end{eqnarray}
where $\eta=PP_1P_2(-1)^{J-J_1-J_2}$ with $P$ and $P_{1,2}$ being the parities and $J$ and $J_{1,2}$ being the angular momenta for the system and particle 1 or 2.
Here  the sum extends only over positive $\lambda''$. The potential with fixed parity is
\begin{eqnarray}
{\cal V}^{J^P}_{\lambda\lambda'}={\cal V}^{J}_{\lambda\lambda'}+\eta {\cal V}^J_{\lambda,-\lambda'},
\end{eqnarray}
which is analogous to the scattering amplitude ${\cal M}^{J^P}_{\lambda\lambda'}$.

Before solving the one-dimensional partial wave BSE numerically,
we need to deal with the pole in $G_0({\rm p})$,
\begin{eqnarray}
	i{\cal M}({\rm p},{\rm p}')&=&i{\cal V}({\rm p},{\rm p}')+\int_0^{{\rm p}''_{max}}\frac{{\rm p}''^2d {\rm p}''}{(2\pi)^3}
	i{\cal V}({\rm p},{\rm p}'') G_0({\rm p}'')i{\cal M}({\rm
	p}'',{\rm p'})\nonumber\\
	&-&i{\cal V}({\rm p},{{\rm p}}_o'')[\int_0^{ {\rm
	p}''_{max}}\frac{d {\rm p}''}{(2\pi)^3}\frac{A({\rm p}''_o)}{
		{\rm p}''^2-{\rm p}''^2_o}
+i\frac{{\rm p}''^2_o\delta\bar{G}_0({\rm p}''_o)}{8\pi^2}]\nonumber\\
&\cdot&i{\cal
M}({\rm p}''_o,{\rm p}')\theta(s-m_1-m_2)\theta({\rm p}''_{max}-{\rm p}''_o),\label{Eq: BSEd}
\end{eqnarray}
where we use the denotions
\begin{eqnarray}
	&&A({\rm p}''_o)\equiv[{\rm p}''^2({\rm p}''^2-{\rm p}''^2_o)G_0({\rm
	p}'')]_{ {\rm p}''\to{\rm p}''_o}=
-\frac{{\rm p}''^2_o}{2W}\nonumber\\
&&\delta \bar{G}_0({\rm p}'')\delta({\rm p}''-{\rm p}''_o)\equiv\delta(G_0(\rm
p)'')=\frac{1}{4W{\rm p}''_o}\delta({\rm p}''-{\rm p}''_o),\quad\quad
\end{eqnarray}
with ${\rm p}''_o=\frac{1}{2W}\sqrt{[W^2-(M+m)^2][W^2-(M-m)^2]}$. Here a cut off of ${\rm p}''$, ${\rm p}''_{max}$, has been considered. The sum of second term  and the first part of third term on the right side of Eq. (\ref{Eq: BSEd}) is the principle value which is calculated with the help of the relation
\begin{eqnarray}
{\cal P}\int^\infty_0 dp\frac{f(p)}{p^2-p_o^2}=\int^\infty_0 dp\frac{f(p)-f(p_o)}{p^2-p_o^2}.
\end{eqnarray}
The $\theta$ functions mean that the pole only appears if $s>m_1+m_2$ and $p''_{max}>p''_o
$.

It is easy to see that the above equation can be related to the formalsim used by Oset $et\ al.$ if the potential kernel
${\cal  V}$ is only dependent on $s$ and $G_0$ is chosen as the same one
used in Ref. \refcite{Oset:1997it}.  In the above derivation the cutoff
regularization is adopted as in Ref. \refcite{Oset:1997it}. In this work we will adopt an  exponential
regularization by introducing a form factor in the propagator as
\begin{eqnarray}
	G_0(p)\to G_0(p)\left[e^{-(k^2-m^2)^2/\Lambda^4}\right]^2,
\end{eqnarray}
and let ${\rm p}''_{max}\to \infty$. Here the baryon is not involved in the form factor due to its
onshellness. The cut off $\Lambda$ plays an analogous role as the cut
off ${\rm p}''_{max}$.

In this work, the covariant spectator theory instead of the one-shell factorization is adopted, so the integrand of the integration about momentum includes not only the propagator but also the potential.  To Solve the integral equation, we discrete the momenta ${\rm p}$,
${\rm p}'$ and ${\rm p}''$ by the Gauss quadrature with weight $w({\rm
p}_i)$ and have,
\begin{eqnarray}
i{M}_{ik}
&=&i{V}_{ik}+\sum_{j=0}^N i{ V}_{ij}G_ji{M}_{jk},
\end{eqnarray}
with the discretized propagator
\begin{eqnarray}
	G_{j>0}&=&\frac{w({\rm p}''_j){\rm p}''^2_j}{(2\pi)^3}G_0({\rm
	p}''_j), \nonumber\\
G_{j=0}&=&-\frac{i{\rm p}''_o}{32\pi^2 W}+\sum_j
\left[\frac{w({\rm p}_j)}{(2\pi)^3}\frac{ {\rm p}''^2_o}
{2W{({\rm p}''^2_j-{\rm p}''^2_o)}}\right].\quad\quad
\end{eqnarray}
In numerical solution, $N$ should be large enough to produce stable result. In the current work $N=50$ is chosen.

In the calculation of the cross section of  certain reaction, the initial and final
particles should be on-shell. The scattering amplitude is
\begin{eqnarray}
	\hat{M}=M_{00}=\sum_j[(1-{ V} G)^{-1}]_{0j}V_{j0}.
\end{eqnarray}
The total cross section can be written as
\begin{eqnarray}
\sigma&=&\frac{1}{16\pi
s}\frac{|{\rm p}'|}{|{\rm p}|}\sum_{J^P,\lambda \geq0
\lambda'\geq0}\frac{{2J+1}}{2}
\left|\frac{\hat{{M}}^{J^P}_{\lambda\lambda'}}{4\pi}\right|^2.
\end{eqnarray}
Note that the second sum extends only over positive $\lambda$ and
$\lambda'$.
Since there is no interference between the contributions from different partial waves, the
total cross section can also be divided into partial-wave cross sections, allowing a direct
access to the importance of the individual partial waves.

\section{Numerical results}

In the current model there are two parameters to be determined, the chiral constant $f$ and the cut off  $\Lambda$ in the
exponential form factor. In this work, we will not make a
fit of the experimental data, so the $f$ for all
channels is chosen as the value used in Ref. \refcite{Oset:1997it},
$f=1.15f_\pi$ with $f_\pi=93$ MeV. As usual
the cut off for channel $l$ can be written as $\Lambda^l=m^l+\alpha \Lambda_{QCD}$ with
meson mass $m^l$, $\Lambda_{QCD}=0.22$ GeV and a common $\alpha$ which is used for all
channels.

\subsection{The cross section for $K^-p\to MB$ reactions}

To determine the only free parameter $\alpha$, the total cross section
for  the $k^-p\to MB$ reactions will be considered.
Since there is only one free parameter, we do not fit the total cross
section directly but find best value of $\alpha$ to reproduce the following threshold ratio,
\begin{eqnarray}
\gamma&=&\frac{\Gamma_{K^-p\to\pi^+\Sigma^-}}{\Gamma_{K^-p\to\pi^-\Sigma^+}}=2.36\pm0.04,\nonumber\\
R_c&=&\frac{\Gamma_{K^-p\to\pi^\pm\Sigma^\mp}}{\Gamma_{K^-p\to{\rm
inelastic}}}=0.664\pm0.011,\nonumber\\
R_n&=&\frac{\Gamma_{K^-p\to\pi^0\Lambda}}{\Gamma_{K^-p\to{\rm
neutral}}}=0.189\pm0.011.
\end{eqnarray}

The best value of $\alpha$ is found at $1.63$, and the results of
the ratios are listed in Table~\ref{Table: ratio}. The scattering
length for $K^-p\to K^-p$ is also listed in the table. The results are comparable with the
experimental values except that the real part of the scattering length $a_{K^-p}$  is larger
than the experimental value as in Ref. \refcite{Oset:1997it}.
\renewcommand\tabcolsep{0.5cm}
\begin{table}[h!]
\caption{The threshold ratios and the scattering
	length.
	\label{Table: ratio}}
\begin{center}
	\begin{tabular}{lrrrr}\hline
	& $\gamma$ & $R_c$ & $R_n$ & $a_{K^-p}$ \\\hline
This work & 2.36 &0.619 & 0.221 & -1.10+0.78 i\\
Ref. \refcite{Oset:1997it} & 2.33 &0.640 & 0.217 & -0.99+0.97 i\\
exp.\cite{Nowak:1978au,Tovee:1971ga} & 2.36 &0.664 & 0.189 &-0.66+0.81 i\\
&$\pm0.04$ & $\pm0.011$ & $\pm0.011$ &$(\pm0.07)+(\pm0.15)$i \\
\hline
\end{tabular}
\end{center}

\end{table}

\begin{figure}[h!]
\begin{center}
\includegraphics[scale=1]{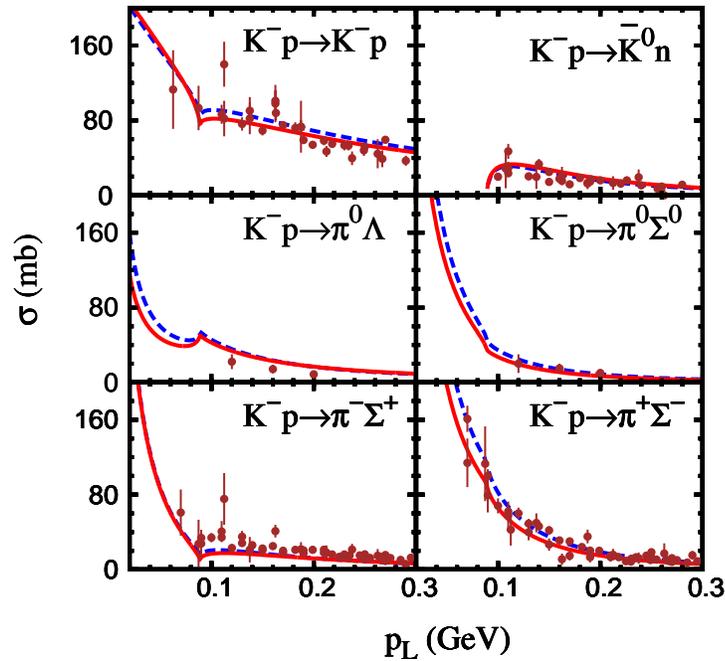}
\caption{Total cross section for $K^-p\to MB$. The solid and dashed lines are for the results in this work and the results of Oset $et.\ al.$ \cite{Oset:1997it}  The experimental data
	are
	from experiments~\cite{Ciborowski:1982et,Humphrey:1962zz,Sakitt:1965kh,Kim:1965zzd,Bangerter:1980px,Evans:1983hz}.}
	\label{Fig: tcs}
\end{center}
\end{figure}

With determined parameters, the total cross section for $K^-p\to
MB$ reactions with $MB=K^-p$, $\bar{K}^0n$, $\pi^0\Lambda$, $\pi^0\Sigma^0$,
$\pi^-\Sigma^+$ and $\pi^+\Sigma^-$ is calculated and shown in Fig.
\ref{Fig: tcs}.

 The experimental data are well reproduced with the
parameters determined by the threshold ratios and close to the results from the off-shell factorization \cite{Oset:1997it}.  The cross section is
dominated by the contribution from the partial wave $J^P=1/2^-$ and the contribution from
the partial wave $J^P=1/2^+$ is negligible and not presented
explicitly in the figure. It suggests that the $P$-wave contribution is negligible compared with the $S$-wave contribution.

\subsection{The poles of the scattering amplitude and the $\Lambda(1405)$}

Now that the model is fixed by the experimental data of the cross
sections, we will search the poles of the scattering amplitude, which are related to the
$\Lambda(1405)$, by extrapolation from the real axis to a complex plane.
The pole can be searched by variation of $z$ to satisfy
\begin{eqnarray}
	|1-V(z)G(z)|=0,
\end{eqnarray}
where  $z=E_R+i\Gamma/2$ equals to the meson-baryon energy $W$ at the real axis. Since
$E=\sqrt{m^2+{\rm p}^2}+\sqrt{M^2+{\rm p}^2}$, the $\rm p$-plane
corresponds to two Reimann sheets for $E$. The bound state is located
in the first Reimann sheet while the resonances located in the second
Reimann sheet with Im(p)$<$0.

\begin{figure}[h!]
\begin{center}
\includegraphics[scale=1]{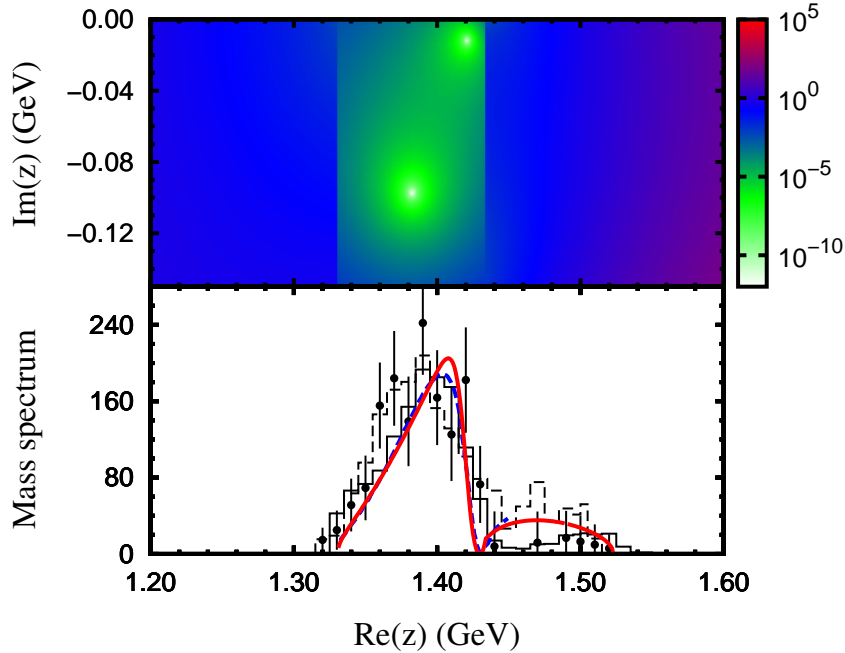}
\caption{The $|1-G(z)V(z)|$ for partial wave $J^P=1/2^-$ in the complex energy plane  and the $\pi\Sigma$ mass spectrum.
The solid, dashed lines are for the result in this work  and
result by Oset and Ramos~\cite{Oset:1997it}. The dashed line is covered by the solid line at some energies. The solid, histogram, dashed histograms and filed dot are for processes $K^- p\to\pi^+\pi^-\Sigma^+\pi^-$~\cite{Hemingway:1984pz} and $\pi^-p\to K^0(\Sigma\pi)^0$~\cite{Thomas:1973uh} and ~$pp\to pK^+(\Lambda(1405)\to \pi^0(\Sigma^0\to \gamma(\Lambda\to p\pi^-)))$~\cite{Zychor:2007gf}. The theoretical results are normalized to the experiment~\cite{Hemingway:1984pz} with solid histogram. } \label{Fig: Swave}
\end{center}
\end{figure}

In Fig. \ref{Fig: Swave} we present the results for the partial wave $J^P=1/2^-$.
Two poles at $1383+99 i$ and $1423+14i$ MeV are produced
from four coupled channels $\bar{K}N$,
$\pi\Sigma$,$\eta\Lambda$ and $K\Xi$.

The invariant mass spectrum of the $\Sigma\pi$ channel is also presented and
compared with the experimental results. The invariant mass
distribution is given approximately as \cite{Hyodo:2003jw}
\begin{eqnarray}
	\frac{d\sigma}{dW}=C|\hat{M}^{J^P}|^2\lambda^\half(W^2,M^2,m^2)\lambda^\half
	(\tilde{W}^2,W^2,m_3^2)/W, \label{Eq: mass}
\end{eqnarray}
where $\tilde{W}$ is the total energy of the process and $m_3$ is the
third final particle in Ref. \refcite{Hemingway:1984pz}, and $M^{J^P}$ is the scattering amplitude for $\Sigma\pi\to\Sigma\pi$. The theoretical result is close to the
experimental data and the peak for the $\Lambda(1405)$ is well
reproduced.  The results are similar to these of Oset and Ramos \cite{Oset:1997it}.

\begin{figure}[h!]
\begin{center}
\includegraphics[scale=1]{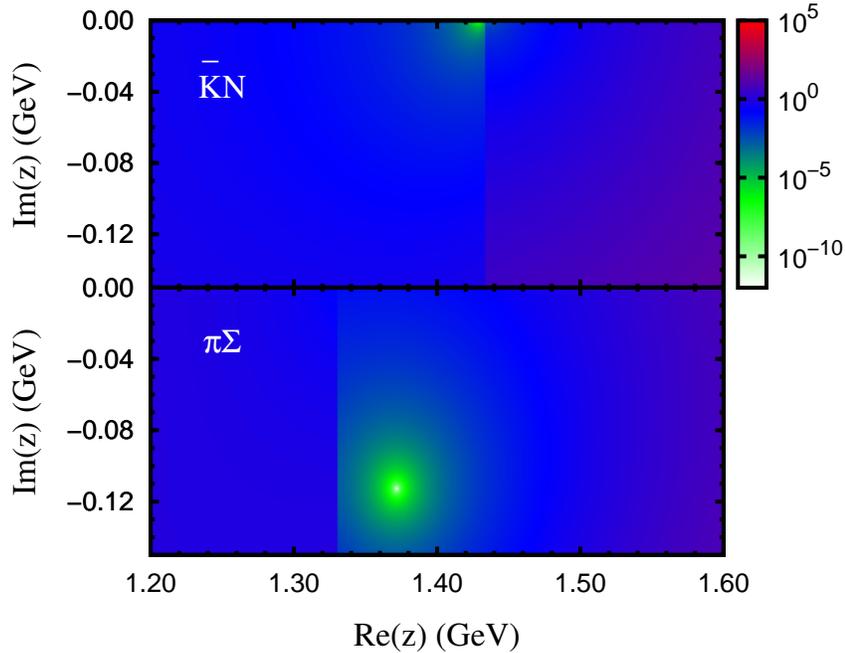}
\caption{The $|1-G(z)V(z)|$ for partial wave $J^P=1/2^-$ in the complex
	energy plane with channel $\bar{K}N$ and $\pi\Sigma$,
respectively. } \label{Fig: 12}
\end{center}
\end{figure}

Since the $\Lambda(1405)$ lies between the $\bar{K}N$ and $\pi\Sigma$
thresholds, it is natural to expect that the two poles originate in
the attractive interactions of these channels. This picture can be
verified by switching off other channels. In Fig. \ref{Fig: 12} we present
the pole positions of the scattering amplitude in the complex energy
plane with  $\bar{K}N$ and $\pi\Sigma$ channels, respectively.
The $\bar{K}N$ channel supports a bound state while a resonance is generated in
the $\pi\Sigma$ channel as shown in Fig. \ref{Fig: 12}.

\begin{figure}[h!]
\begin{center}
\includegraphics[scale=1]{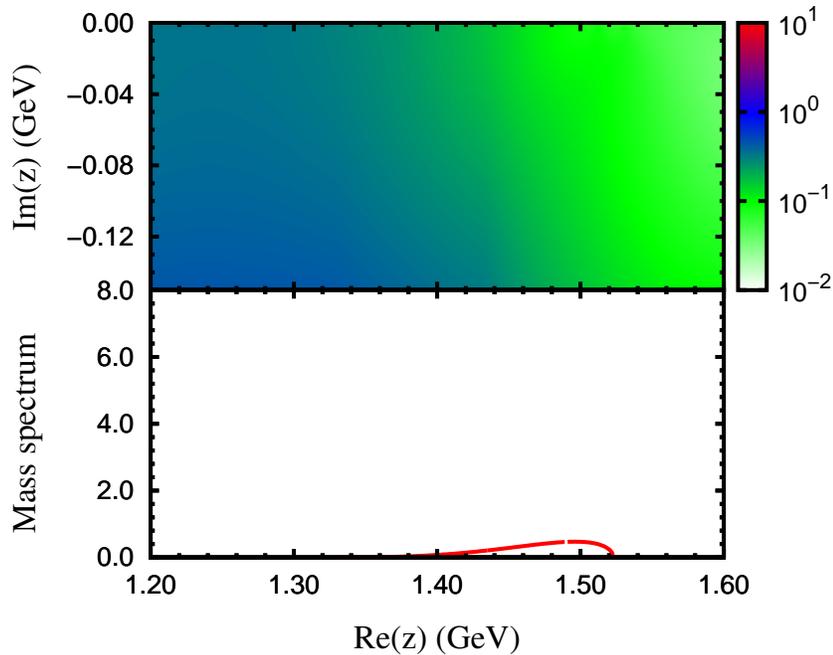}
\caption{The $|1-G(z)V(z)|$ for partial wave $J^P=1/2^+$ in the complex energy plane  and the $\pi\Sigma$ mass spectrum.} \label{Fig: Pwave}
\end{center}
\end{figure}

The scattering amplitude with $J^P=1/2^-$ is a S-wave contribution.
The potential also provides a P-wave contribution with $J^P=1/2^+$
which is proportional to the $pp'$ as shown in Eq. (\ref{Eq:
potential with d}). The results for the total cross section of $K^-p\to
MB$ reactions suggest its contribution is negligible. In Fig. \ref{Fig: Pwave},
we present the  pole positions of  scattering amplitude  for partial wave $J^P=1/2^+$ in the
complex energy plane  and the $\pi\Sigma$ mass spectrum.  One can find that the potential
for $J^P=1/2^+$ is very weak so that no pole is produced. The results is consistent with  Ref. \refcite{Jido:2002zk} where the $p$-wave contribution was studied with  unitarized  amplitudes $f_+$ and $f_-$ instead of the helicity amplitudes used in this work and also found to be small.

\section{Summary and Conclusions}

The octet meson and octet  baryon interaction with strangeness and its relation to the $\Lambda(1405)$ are
studied in a Bethe-Salpeter
equation approach with quasipotential approximation. In this work the covariant spectator theory instead of the on-shell factorization are adopted and the integral equation about the helicity amplitude is solved after partial wave expansion.
The total cross section for $K^-p\to MB$ reactions with $MB=K^-p$, $\bar{K}^0n$, $\pi^-\Sigma^+$, $\pi^0\Sigma^0$,
$\pi^+\Sigma^-$ and $\eta \Lambda$ are calculated with all possible partial
waves and compared with the experimental data. Two poles at $1383+199 i$ and $1423+14i$ MeV are
produced, which originate from the $\pi\Sigma$ interaction as a resonance and the $\bar{K}N$ interaction as a bound state, respectively. In this work, only the Weinberg-Tomozawa term is considered in the calculation. The experimental data can be reproduced generally and close to the results in Ref.  \refcite{Oset:1997it}. However, there still exist considerable discrepancies between the current model and experiment.  To give a better description of the experimental data, a more comprehensive investigation, such as explicitly fitting and including $s$ and $u$ channels~\cite{Oller:2005ig,Borasoy:2004kk,Oller:2006jw,Ikeda:2011pi} and NLO terms, in our approach is needed.

\section*{Acknowledgements}
This project is partly supported by the Major
State Basic Research Development Program in China
under grant 2014CB845405,
the National Natural Science Foundation of China under Grant 11275235.

\bibliographystyle{elsarticle-num}

\end{document}